# Dissociation energies of Ag–RG (RG = Ar, Kr, Xe) and AgO molecules from velocity map imaging studies

Graham A. Cooper,[1] Aras Kartouzian,[1,2] Alexander S. Gentleman,[1] Andreas Iskra,[1] Robert van Wijk,[1] and Stuart R. Mackenzie[1]*

[1]Oxford Chemistry, Physical and Theoretical Chemistry Laboratory, South Parks Road, Oxford, OX1 3QZ, United Kingdom

[2]Technische Universität München, Department of Physical Chemistry, Catalysis Research Center, Lichtenbergstr. 4, 85748 Garching, Germany

KEYWORDS: velocity map imaging, metal cluster, dissociation energy, photodissociation

ABSTRACT

The near ultraviolet photodissociation dynamics of silver atom – rare gas dimers have been studied by velocity map imaging. Ag–RG (RG = Ar, Kr, Xe) species generated by laser ablation are excited in the region of the $C\ (^2\Sigma^+) \leftarrow X\ (^2\Sigma^+)$ continuum leading to direct, near-threshold dissociation generating Ag* ($^2P_{3/2}$) + RG ($^1S_0$) products. Images recorded at excitation wavelengths throughout the $C\ (^2\Sigma^+) \leftarrow X\ (^2\Sigma^+)$ continuum, coupled with known atomic energy levels, permit determination of the ground $X\ (^2\Sigma^+)$ state dissociation energies of 85.9 ± 23.4 cm$^{-1}$ (Ag–Ar), 149.3 ± 22.4 cm$^{-1}$ (Ag–Kr) and 256.3 ± 16.0 cm$^{-1}$ (Ag–Xe). Three additional photolysis processes, each yielding Ag atom photoproducts, are observed in the same spectral region. Two of these are markedly enhanced in intensity upon seeding the molecular beam with nitrous oxide, and are assigned to photodissociation of AgO at the two-photon level. These features yield an improved ground state dissociation energy for AgO of 15965 ± 81 cm$^{-1}$, which is in good agreement with high level calculations. The third process results in Ag atom fragments whose kinetic energy shows anomalously weak photon energy dependence and is assigned tentatively to dissociative ionization of the silver dimer Ag$_2$.

## I. Introduction

Neutral coinage metal atom – rare gas dimers represent a type of van der Waals molecule bound in the ground state predominantly by dispersion interactions. A better understanding of the nature of these fundamental interactions would inform fields as diverse as molecule surface interactions[1] and buffer gas cooling.[2] The study of such molecules has been lent added impetus in recent years, by the popular use of rare gas atoms as messenger tags in action spectroscopy studies of gas-phase metal–ligand[3, 4] naked metal[5-8] and metal oxide clusters.[9-11] In such studies, the structures of interest are widely assumed to be affected minimally by the presence of the inert tag with the result that the photodissociation action spectrum mirrors that of the naked species. In fact, the role of the tag can be more subtle and there are clear examples in which clustering with rare gas atoms changes the distribution of low-lying structures.[12]

On the whole, spectroscopic studies of metal atom – rare gas molecules have focused on regions around strong atomic transitions. In the case of the silver atom with its $[Kr]4d^{10}5s^1$, alkali metal–like configuration, this is the fully allowed $5p \leftarrow 5s$ transition. Figure 1 shows the schematic potential curves for the electronic states of Ag–RG relevant for the present study. The curves shown are for Ag–Ar and represent those states explored experimentally by Jouvet *et al.*[13] and by Brock and Duncan.[14] The same states have recently been the subject of a detailed computational study at the spin-unrestricted coupled cluster level (UCCSD(T)) by Loreau *et al.*.[15] A different excited state labelling is employed by Loreau *et al.* but we have adopted labels consistent with Brock and Duncan. All states of interest correlate with ground state Ar atoms plus low–lying $4d^{10}5s$ ($^2S$), $4d^{10}5p$ ($^2P$) and $4d^95s^2$ ($^2D$) terms of Ag. Qualitatively similar curves exist for the other Ag–RG complexes[15] although Loreau *et al.* recently showed that the excited state potentials show marked deviation from Morse functions when spin-orbit coupling is included. The *A* states of Ag–He and Ag–Ne are particularly anomalous supporting double well structures.

Jouvet *et al.* studied Ag-Ar *via* fluorescence excitation of the $A(^2\Pi_i)$–*X* and $B(^2\Pi_{3/2})$–*X* transitions and determined excited state dissociation energies by Birge–Sponer extrapolation.[13] The same transitions were subsequently studied by Brock and Duncan in a wider range of Ag–RG (RG=Ar, Kr, Xe) dimers by resonant two-photon ionization (R2PI).[14] Again, Birge–Sponer and/or LeRoy–Bernstein[16, 17] extrapolations were used to estimate the excited state dissociation thresholds which were then coupled with the well-known atomic energy levels permitting determination of the ground state dissociation energies.

We have previously applied the velocity map imaging (VMI)[18, 19] technique to the study of the photodissociation dynamics of a range of neutral cluster species including $Au_2$,[20] Au–RG (RG=Ar, Kr, Xe),[21, 22] $Li(NH_3)_4$,[23] $Xe_2$,[24, 25] $Cu_2$ and CuO.[26] These studies have demonstrated the power of VMI in revealing subtle details of the interactions between weakly–bound neutral species, especially those containing metal atoms. In many cases, these studies have led to improved dissociation energies for the molecules studied. Here, we extend these studies to small silver atom – containing molecules for comparison with the other coinage metal molecules.

In optimizing the production of Ag–RG species in our instrument, it was useful to reproduce the R2PI spectra of Ag–Ar, Ag–Kr and Ag–Xe *via* the $A$ ($^2\Pi_i$) states reported by Brock and Duncan but we cannot improve on their spectra. In these studies, we have explored the near-threshold dissociation of these complexes *via* excitation to the repulsive wall of the extremely weakly–bound $C$ ($^2\Sigma^+$) state. In doing so, we have applied VMI to measure directly the kinetic energy released in the fragmentation process and thereby better determine the dissociation threshold and, hence, the ground state dissociation energy.

In this work we report VMI studies of a range of neutral Ag–X (X=Ar, Kr, Xe, O, Ag) molecules generated by entraining silver atoms generated by laser ablation in a seeded backing gas prior to supersonic expansion into vacuum. This method presents a number of unique problems. In most VMI photodissociation studies the identity of the precursor species is clear - usually either a stable precursor molecule or a mass–selected ion.[19] Alternatively VMI can be used for detecting the product state distributions of chemical reactions of neutral[19, 27, 28] or ion-molecule reactions.[29-31] In our experiments, however, by virtue of the laser ablation technique, a range of species is often present within our molecular beam and much of the analysis involves identifying the co-fragment of a photodissociation event, and hence the identity of the parent molecule. The situation is further complicated by multiple-photon processes and the possibility of dissociation on both neutral and ionic potential energy surfaces.

At heart, photodissociation VMI relies on conservation of energy, which, for the prototypical Ag–X species studied here can be represented as

$$E_{AgX} + nh\nu = D_0(Ag - X) + E_{Ag} + E_X + TKER, \qquad (1)$$

in which $D_0(Ag - X)$ is the experimental dissociation energy; $E_{AgX}$, $E_{Ag}$ and $E_X$ are the internal energies of the precursor molecule, the Ag fragment and the X fragment, respectively, and TKER is the total kinetic energy release. The momenta of the two fragments are equal and opposite in the centre of mass frame, and thus measurement of the kinetic energy distribution for one fragment is sufficient to determine the full TKER distribution provided the mass of the co-fragment is known. In the experiments described here it proved convenient to measure the kinetic energy distribution of the Ag atom fragment which, for a given dissociation channel, is given by

$$KE_{Ag} = \frac{p_{Ag}^2}{2m_{Ag}} = \frac{m_X}{m_{AgX}}.TKER \propto nh\nu . \qquad (2)$$

Combining this with eqn (1) it is clear that, where dissociation into the same channel occurs across a range of photon energies, $h\nu$, the slope of graph of $KE_{Ag}$ as a function of photon energy yields the mass of the cofragment and the number of photons involved in the photodissociation step. Similarly, the *x*-intercept (*TKER* = 0) provides information on the energy of the dissociation threshold involved, from which the ground state dissociation energy may be determined.

The details of the instrumentation and experimental methodology are given in Section II. The results and discussion in Section III are separated into discussion of the Ag-RG complexes, AgO and $Ag_2$ as the processes involved are qualitatively different. Section IV summarises the major findings of these studies.

## II. Experimental

The laser ablation VMI instrument used in these studies has been described in detail previously.[20, 21, 26] Briefly, silver atoms are generated by laser ablation of a rotating silver metal target and entrained in a pulsed molecular beam of either pure or mixed rare gases delivered via a Jordan valve. For the studies of Ag-Ar pure argon carrier gas was used. To generate Ag–Kr (Ag–Xe), 10% Kr (Xe) in argon was used. Photofragmentation / photoionization occurs within a VMI spectrometer arranged with ion extraction axis collinear with the cluster beam. Photoexcitation in the region 345 – 315 nm was provided by the frequency doubled output of a pulsed dye laser pumped by the 2$^{nd}$ harmonic of a Nd:YAG laser. Both $^{107}$Ag$^+$ and $^{109}$Ag$^+$ are detected in proportion to their naturally occurring abundance, and gating over a single isotope slightly increases the resolution of the VMI but only at the expense of proportionately lower count rate. The images shown here were gated over both silver isotopes and were recorded at 10 Hz for typically up to an hour. The instrument is not configured for slice imaging[19, 32] and image reconstruction was performed using a polar onion peeling algorithm developed by Roberts *et al.*.[33] We also used the maximum entropy method of Dick[34] which we find to perform more reliably especially at low signal intensities or in identifying minor dissociation channels.

Production of Ag-RG dimers was optimised by reproducing the *A* ($^2\Pi$) ← *X* R2PI spectra of Brock and Duncan. Unlike in the case of Au–RG dimers we have studied previously,[21] no dissociation of the bound *A* ($^2\Pi$) state levels was observed.

As detailed in Section IIIB below, some features observed in the VMI images were tentatively assigned as originating from the AgO molecule. To test this assignment, *in situ* production of the oxide was enhanced by seeding the helium carrier gas with *ca*. 10% nitrous oxide.

## III. Results and Discussion

A.   Ag–RG (RG = Ar, Kr, Xe)

Having confirmed the production of Ag–Ar dimers by means of R2PI spectroscopy,[14] the laser is tuned into the *C* ← *X* continuum at wavelengths λ < 327 nm. By remarkable coincidence, the Ag* $^2P_{3/2}$ state with which the *C* ($^2\Sigma^+$) state correlates can be ionized by single photon excitation[35] at wavenumbers > 30634 cm$^{-1}$ ($\lambda$ < 326.4 nm) and thus Ag$^+$ is formed efficiently in a combined single-colour photodissociation / ionization process:

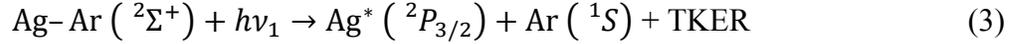

$$\text{Ag–Ar}\,(\,^2\Sigma^+) + h\nu_1 \rightarrow \text{Ag}^*\,(\,^2P_{3/2}) + \text{Ar}\,(\,^1S) + \text{TKER} \qquad (3)$$

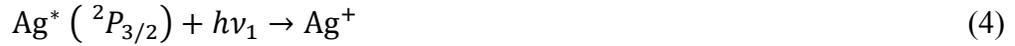

$$\text{Ag}^*\,(\,^2P_{3/2}) + h\nu_1 \rightarrow \text{Ag}^+ \qquad (4)$$

Figure 2 shows typical VMI images observed in this region by gating the MCP detector over the Ag$^+$ signal. At these excitation wavenumbers, an intense central spot, attributed to a combination of direct ionization of metastable Ag* and two–photon ionization of ground state Ag within the cluster beam is observed and has been omitted from the images and spectra in Figure 2 for clarity. The left hand side of the image shows the raw image data and the right hand side a slice through the reconstructed 3D distribution following onion peeling.

The VMI images are dominated by a single intense ring, the radius of which increases with increasing laser wavenumber. The images show a marked but not extreme parallel character (the best fit $\beta_2$ anisotropy parameter is *ca.* +1.6 ± 0.1) consistent with the Σ–Σ transition excited. It is important to note that although the images are shown on the same scale they represent only the central 10% area of the detector (hence the pixellation of the images shown). These are near-threshold dissociation processes with very low kinetic energy release. The spectra shown in Figure 2 illustrate how the Ag* kinetic energy distribution changes as a function of excitation energy. These distributions are derived directly from the momentum space images and corrected for the low magnification factor (*ca.* 1.06) of our shaped electrode arrangement.

Figure 3 summarises all the dissociation channels observed in this study including the data shown in Figure 2. The Ag–Ar data is shown at photon energies above 30500 cm$^{-1}$ together with the equivalent data obtained in studies of Ag–Kr and Ag–Xe. Each dissociation channel observed (each ring in a VMI image resulting in a clear peak in the extracted kinetic energy release spectrum) is represented by a point on the graph. The uncertainties quoted reflect the full width at half maximum of the corresponding peak in the kinetic energy release spectrum – a conservative measure of how well the peak centre is determined. For these small rings in the VMI images, this uncertainty can be up to *ca*. 10% of the kinetic energy release itself and is limited by a combination of molecular beam temperature, pixel resolution and, in some cases, the unresolved isotope distribution of the rare gas co-fragment.

The slopes of the data sets in Figure 3 ($= n(m_X/m_{AgX}).TKER$, see eqn (2)), corresponding to the dissociation of Ag–RG species, reflect the differing kinetic energy the Ag* fragment assumes upon recoil from different mass co-fragments and confirm the single–photon nature of the dissociation process involved. Extrapolation of these fits to zero kinetic energy release provides a precise and accurate measure of the relevant Ag ($^2P_{3/2}$) + RG ($^1S$) dissociation threshold relative to the zero point level of the Ag–RG ground state out of which excitation is assumed to occur. Given that the atomic energy levels of the silver atom are well known, the ground state dissociation energies may be determined for comparison with the results of earlier studies. Such a comparison is given in Table 1.

**Table 1**: Ground state experimental dissociation energies, $D_0$ for the Ag-RG dimers

|  | $D_0$ /cm$^{-1}$ (this work) | Calc. /cm$^{-1}$ | Expt.[c] /cm$^{-1}$ | Expt.[d] /cm$^{-1}$ |
|---|---|---|---|---|
| Ag–Ar | 85.9 ± 23.4 | 102.62[a]<br>104.2[b] | 65<br>51 | 77 ± 15<br>35 ± 45 |
| Ag–Kr | 149.3 ± 22.4 | 163.61[a]<br>160.3[b] | 138<br>68 | |
| Ag-Xe | 256.3 ± 16.0 | 254.25[a]<br>244.6[b] | 276<br>179 | |

[a] CCSD(T) Loreau *et al.*[15]
[b] RCCSD(T), Gardner *et al.*[36]
[c] Brock and Duncan[14] values from Birge-Sponer fits to *A-X* data for $^{107}$Ag$^{40}$Ar, $^{107}$Ag$^{83}$Kr, $^{107}$Ag$^{129}$Xe
[d] Jouvet *et al.*[13] – the different values come from Birge-Sponer extrapolation of *A-X* data.

The key advantage that the current measurements have over previous experimental determinations is that they do not rely upon extrapolation of vibrational progressions which was the source of the variation in previous measurements. Brock and Duncan performed Birge-Sponer and LeRoy-Bernstein fits to their high–quality resonant photoionization spectra of the *A* $^2\Pi_{3/2,1,2}$; *B* $^2\Pi_{3/2}$ ← *X* transitions. Different data sets, however, led to markedly different values when extrapolated to threshold. Each extrapolation method relies on a particular form (*e.g.*, Morse, -$R^m$) for the potential energy curve and it is now clear that the excited states of Ag–RG atoms do not follow such simple forms. Indeed Loreau *et al.* have suggested the deviation from Morse behavior in the case of the *A* $^2\Pi_{1/2}$ state of Ag–Ar is so marked that the vibrational assignment of the spectra is incorrect.[15] Agreement of the current results with calculated values is good with the computational $D_0$ values lying well within the experimental uncertainties.

Comparison of the $D_0$ values for the Ag-RG dimers (Table 1) with those previously obtained by this group for the analogous Au-RG species(149 ± 13 cm$^{-1}$ (AuAr), 240 ± 19 cm$^{-1}$ (AuKr) and 607 ± 5 cm$^{-1}$ (AuXe)),[21] suggests a contrast in the nature of the bonding in each series. The dissociation energies of the Ag-RG species (Table 1) scale linearly with the static dipole polarizabilities of the rare gas atoms (11.1, 16.7 and 27.3 atomic units for Ar, Kr and Xe respectively) reflecting the predominant dispersion nature of the interaction.[37] Those of the Au-RG species, however, show a marked greater than linear dependence indicating the increased role of non-dispersion interactions. This is particularly clear in the much larger dissociation energy of AuXe compared to both AuKr and AgXe.[22]

**B.  Other dissociation processes: 2-photon dissociation of AgO**

During the measurements on Ag-RG species, we took the opportunity to look for evidence of predissociation of the bound *A*, *B* state levels in the region 29000 – 30500 cm$^{-1}$ particularly for Ag–Ar. As shown in Figure 3, the signatures of three additional dissociation processes were indeed observed. These, however, are much weaker than the Ag–RG channels described above and proved more challenging to assign. Two of the features, those corresponding to $KE_{Ag*} \approx 200$ cm$^{-1}$ and 1200 cm$^{-1}$, show a photon energy dependence to the $KE_{Ag*}$ which apparently matches that for the one–photon Ag–Ar dissociation considered above suggesting it is the carrier. However, the observation of the rings shows no correlation with individual vibrational bands in the $A\ ^2\Pi_{3/2} \leftarrow X$ R2PI spectrum of Ag–Ar with the rings in the VMI images equally intense between spectral features as on resonance.

Instead, we assign these two features to two-photon dissociation of AgO. The full dataset includes eight independent measurements of the TKER separation for the two channels involved. Combined, these produce an energy separation of 7721 ± 191.4 cm$^{-1}$ which matches well the separation in energy of the Ag* $4d^9 5s^2\ ^2D_{3/2}$ and $4d^{10} 6s^1\ ^2S_{1/2}$ states (= 7841.8 cm$^{-1}$). We thus assign these channels to the two-photon processes

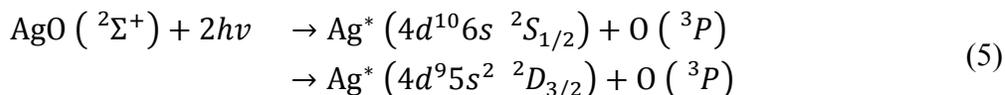

$$\begin{aligned}
\text{AgO}\,(^2\Sigma^+) + 2h\nu &\rightarrow \text{Ag}^*\,(4d^{10}6s\ ^2S_{1/2}) + \text{O}\,(^3P) \\
&\rightarrow \text{Ag}^*\,(4d^9 5s^2\ ^2D_{3/2}) + \text{O}\,(^3P)
\end{aligned} \quad (5)$$

The apparently similar slope of the $KE_{Ag*}(h\nu)$ graph to that for Ag–Ar is coincidental with two-photon dissociation with a $^{16}$O co-fragment giving very similar slope ($2m_O/m_{AgO} = 0.26$) to that for a one-photon process with Ar cofragment ($m_{Ar}/m_{AgAr} = 0.27$).

Confirmation of the assignment of a transition in AgO is provided by the fact that i) the same pair of rings are observed in a pure helium expansion (*i.e.*, in the absence of Ar) and ii) the rings are markedly enhanced upon seeding the molecular beam with nitrous oxide (see Figure 4). $N_2O$ can both react with Ag atoms / clusters within the molecular beam generating gas phase AgO, as well as oxidise the silver surface resulting in AgO ablation.[38] The net result is more AgO in the cluster beam and an increase in the intensity of any rings in the VMI images arising from AgO dissociation.

Having assigned the dissociation channels, simultaneous fitting and extrapolation of both datasets locates the two AgO dissociation thresholds and, together with the known Ag* atomic energy levels, permits determination of the ground state dissociation energy as $D_0 = 15965 \pm 81$ cm$^{-1}$ (see Table 2). This represents an important number as the dissociation energy of AgO has proven challenging to measure experimentally despite its importance in the determination of other parameters.[39] The currently accepted value is an early mass spectrometric determination of $52.7 \pm 3.5$ kcal mol$^{-1}$ (= $2.29 \pm 0.15$ eV ≡ $18400 \pm 1200$ cm$^{-1}$) by Smoes *et al.* using a Knudsen cell.[40] This value has long been considered unreliable, however, representing a significant outlier in comparison with other experimental and calculated values.[41-45]

**Table 2**: Ground state experimental dissociation energy, $D_0$, for AgO

|     | $D_0$ /cm$^{-1}$ (this work) | Smoes *et al.* [40] / kcal mol$^{-1}$ | Calc. |
| --- | --- | --- | --- |
| AgO | $15965 \pm 81$ <br> $16007 \pm 139^a$ <br> $15942 \pm 100^b$ | $52.7 \pm 3.5$ <br> ($18400 \pm 1200$ cm$^{-1}$) | 1.91 eV [46]  (15400 cm$^{-1}$) <br> 1.82 eV [43]  (14700 cm$^{-1}$) <br> 1.71 – 2.2 eV [42]  (13800 – 18000 cm$^{-1}$) <br> 1.74 eV [45]  (14000 cm$^{-1}$) |

$^a$ derived from the Ag* $^2D_{3/2}$ + O $^3P$ channel data in isolation
$^b$ derived from the Ag* $^2S_{1/2}$ + O $^3P$ channel data in isolation

The values determined in this study mark a significant revision in the AgO dissociation energy more in line with, but still marginally higher than, high–level calculations.

Both VMI rings associated with AgO photodissociation exhibit weak perpendicular anisotropy which is markedly less pronounced than that observed in the Ag–RG dissociation, reflecting the two-photon nature of the dissociation (and the role of higher order anisotropy terms such as $\beta_4$). The rings are also slightly broader than those arising from Ag–RG dissociation which may in part arise from the unresolved spin-orbit levels of the atomic O ($^3P$) co-fragments.

### C.  Possible dissociative ionization of $Ag_2$

One, very low kinetic energy release ($KE_{Ag^*}$< 100 cm$^{-1}$) process is observed in images across this whole spectral region. It becomes most prominent, however, in images recorded at photon energies below the two–photon ionization threshold (30550 cm$^{-1}$)[35] where the intense central spot arising from direct ionization of Ag atoms in the cluster beam is reduced. The corresponding peak in the kinetic energy release spectrum is clearly visible in the Figure 4 at $KE_{Ag^*} \approx$ 85 cm$^{-1}$ and as a small ring in the centre of the image b) in the same Figure. Although the kinetic energy release in this channel is very small, a well-defined off–axis velocity component is clearly observed confirming that the detected Ag$^+$ originate from a dissociation process of some description. However, this channel is unusual in that the measured photofragment kinetic energy shows a very weak dependence on photon wavenumber. Over the 500 cm$^{-1}$ photon energy range measured in detail, the Ag fragment kinetic energy varies by less than 10 cm$^{-1}$. Assignment as a two-body dissociation would require recoil from an extremely light co–fragment (mass ~ 2u). The lightest possible cofragment in our experiments would be helium and we have never seen evidence of neutral metal atom–He complexes in any of our experiments to date. Furthermore, the same features in the VMI image are observed in a pure Argon expansion. In principle AgH could be formed by reactions with trace background water but the trends in kinetic energy release with photon energy are inconsistent with this.

Instead, we believe the origin of these low KE Ag atoms lies in dissociative photoionization. In such a process, much of the excess energy available is removed by the departing electron with the remainder partitioned into the atomic / molecular photofragments. Although such channels may represent comparatively unusual processes, they are detected in VMI with high efficiency as the need for ionization of a fragment is obviated. As a result, dissociative photoionization has been observed by VMI in a wide range of systems including diatomic molecules (e.g., $D_2$,[47] $Br_2$,[48] $I_2$[49]), $NO_2$,[50] $CH_3I$,[51] glycidol clusters,[52] ionic liquids,[53] small hydrocarbons,[54] hydrocarbon radicals,[55] rare–gas dimers[25, 56] and Au–RG complexes.[22]

Clues to the identity of the precursor molecule for this process come from Figure 4. In a pure helium expansion, the low KE Ag fragments represent the dominant fraction. However, this channel is greatly depleted upon the addition of nitrous oxide to the beam. The enhanced AgO signal is parasitic of the molecular carrier for the low KER process. The most likely candidate would seem to be small Ag clusters ($Ag_2$, $Ag_3$, etc) which are produced in the cluster beam and which can be observed by direct 157nm photoionization. Such clusters would react with any added $N_2O$ to form AgO consistent with the results in Figure 4. $Ag_2$ has an ionization energy of 61660 cm$^{-1}$ known from zero kinetic energy photoelectron spectroscopy[57] and the $Ag_2^+$ $X\,^2\Sigma_g^+$ ground state a dissociation energy of 12 666 ± 250 cm$^{-1}$.[58] $Ag_2^+$ itself has a well–studied dissociative absorption band at around 3.0 eV[59] and so dissociative ionization at the three photon level is plausible. In such a process, however, without knowledge of the photoelectron energy, it is impossible to infer the quantum states of the atomic fragments and thus the exact dissociation channel.

*IV. Conclusions*

A range of dissociation / ionization processes in neutral silver atom containing molecules / complexes has been explored by UV velocity map imaging detecting $Ag^+$ ions. In all cases, both the number of photons involved in the dissociation and the nature of the co-fragment have been identified by studying the variation in the kinetic energy release with photon energy. In the near–threshold region studied, a rich array of dynamical processes have been observed including direct single–photon dissociation of Ag–RG complexes, two–photon dissociation of AgO and dissociative ionization of $Ag_2$. Analysis of the variation in the kinetic energy release as a function of photon wavenumber permits determination of improved ground state dissociation energies for the Ag–RG (RG = Ar, Kr, Xe) and AgO molecules.

FIGURES:

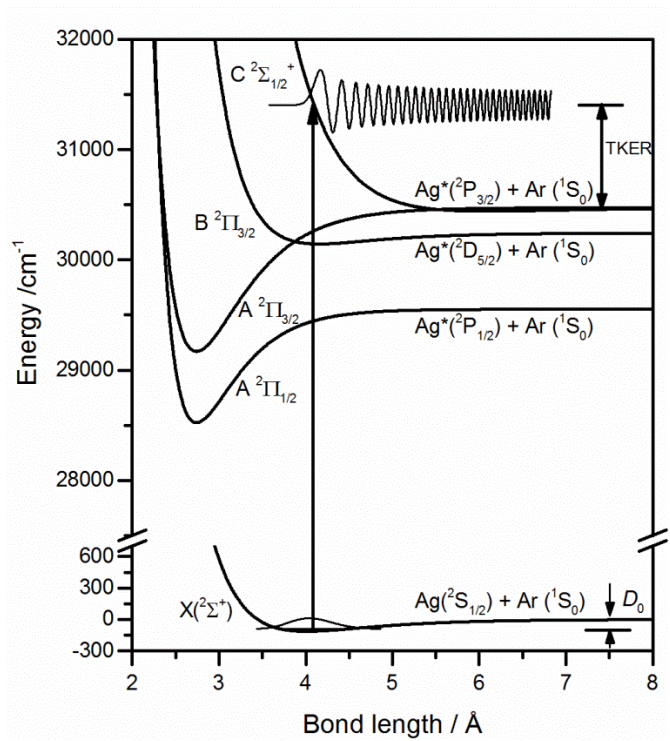

**Figure 1.** Schematic representation of the ground and low-lying excited electronic states of the Ag–Ar molecule.[14, 15] The relevant potential energy curves of Ag–Kr and Ag–Xe are qualitatively similar.

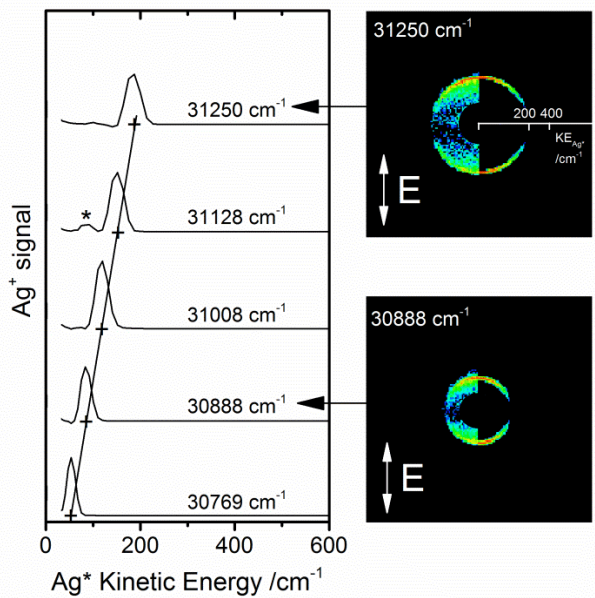

**Figure 2.** Ag* atom photofragment kinetic energy release spectra recorded at different points within the C–X continuum. Representative VMI images recorded at 30888 cm$^{-1}$ and 31250 cm$^{-1}$ are shown to illustrate the larger kinetic energy release in the latter. The left hand side of the images shows the symmetrized raw data, the right hand side central slice of the Newton Sphere reconstructed using the polar onion peeling approach. Both images show clear anisotropy reflecting a parallel transition. The asterisk indicates a weak feature assigned to dissociative ionization of Ag$_2$ (see Section III.C)

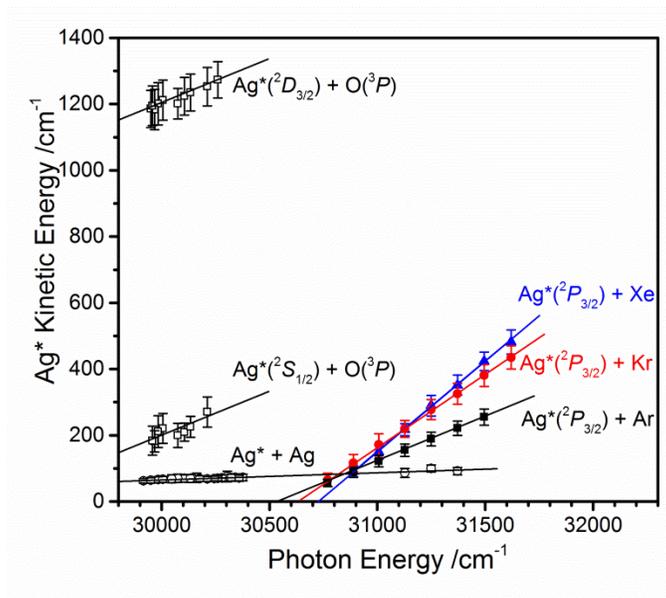

**Figure 3.** Summary of all dissociation processes observed by velocity map imaging. Above 30500 cm$^{-1}$ the dominant process observed is photodissociation of Ag–RG dimers following single-photon excitation in the C-X band. Weaker rings observed at lower excitation energies are assigned to two–photon photodissociation of AgO and dissociative ionization of Ag$_2$ (see text for details).

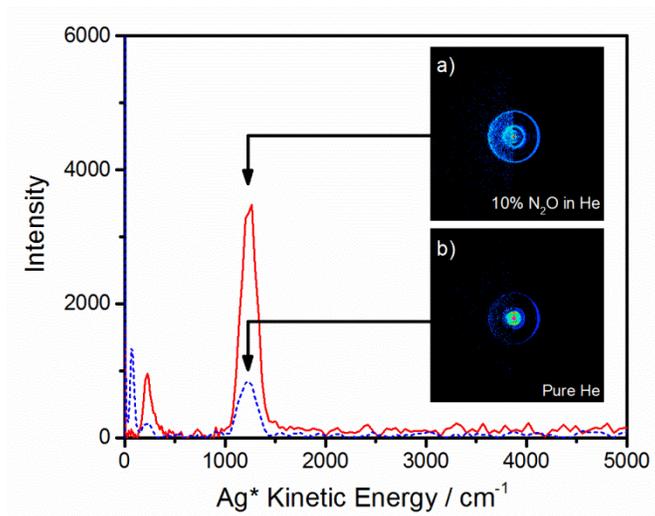

**Figure 4.** VMI images and extracted kinetic energy release spectra following photodissociation of AgO. Seeding the molecular beam with $N_2O$ (solid, red line) significantly enhances the AgO dissociation rings relative to the inner (lowest TKER) ring assigned to dissociative ionization of the $Ag_2$ dimer (see text).

AUTHOR INFORMATION

**Corresponding Author**

*stuart.mackenzie@chem.ox.ac.uk

ACKNOWLEDGMENT

**Funding Sources**

SRM gratefully acknowledges EPSRC for funding (Programme Grant EP/L005913, on which AG is employed). AK is grateful to the Royal Society for his Newton Fellowship which continues to fund his collaboration with Oxford. GAC and AI further thank EPSRC and Magdalen and Wadham Colleges, respectively for their graduate studentships.